\RequirePackage[l2tabu, orthodox]{nag}
\documentclass[%
 aip,
 jmp,%
 amsmath,amssymb,
 reprint,%
]{revtex4-1}
\usepackage{graphicx}
\usepackage{dcolumn}
\usepackage{bm}

\usepackage{wrapfig}
\usepackage{epstopdf}
\usepackage{amsmath}              
\usepackage{amssymb} 
\usepackage{textcomp}
\usepackage{tabularx} 
\usepackage{multirow}
\usepackage[ngerman, english]{babel}
\usepackage{hyperref}

\begin{document}

\preprint{AIP/123-QED}

\title{Versatile fluoride substrates for Fe-based superconducting thin films}

\author{F.~Kurth}
\email[Electronic address:\,]{Fritz.Kurth@ifw-dresden.de}
\affiliation{IFW Dresden, Helmholtzstr.\ 20, 01069 Dresden, Germany}
\affiliation{TU Dresden, 01062 Dresden, Germany}
\author{E.~Reich}
\author{J.~H\"anisch}
\affiliation{IFW Dresden, Helmholtzstr.\ 20, 01069 Dresden, Germany}
\author{A.~Ichinose}
\author{I.~Tsukada}
\affiliation{Central Research Institute of Electric Power Industry, 2-6-1 Nagasaka, Yokosuka, Kanagawa 240-0196, Japan}
\author{R.~H\"uhne}
\affiliation{IFW Dresden, Helmholtzstr.\ 20, 01069 Dresden, Germany}
\author{S.~Trommler}
\affiliation{IFW Dresden, Helmholtzstr.\ 20, 01069 Dresden, Germany}
\affiliation{TU Dresden, 01062 Dresden, Germany}
\author{J.~Engelmann}
\affiliation{IFW Dresden, Helmholtzstr.\ 20, 01069 Dresden, Germany}
\affiliation{TU Dresden, 01062 Dresden, Germany}
\author{L.~Schultz}
\affiliation{IFW Dresden, Helmholtzstr.\ 20, 01069 Dresden, Germany}
\affiliation{TU Dresden, 01062 Dresden, Germany}
\author{B.~Holzapfel}
\affiliation{IFW Dresden, Helmholtzstr.\ 20, 01069 Dresden, Germany}
\affiliation{TU Bergakademie Freiberg, Akademiestr.\ 6, 09596 Freiberg, Germany}
\author{K.~Iida}
\affiliation{IFW Dresden, Helmholtzstr.\ 20, 01069 Dresden, Germany}

\date{\today}

\begin{abstract}
We demonstrate the growth of Co-doped BaFe$_2$As$_2$ (Ba-122) thin films on {\textit AE}F$_2$ (001) ({\textit AE}:Ca, Sr, Ba) single crystal substrates using pulsed laser deposition. All films are grown epitaxially despite of a large misfit of -10.6~\% for BaF$_2$ substrate. For all films a reaction layer is formed at the interface confirmed by X-ray diffraction and by transmission electron microscopy. The superconducting transition temperature of the film on CaF$_2$ is around 27~K, whereas the corresponding values of the other films are around 21~K. The Ba-122 on CaF$_2$ shows identical crystalline quality and superconducting properties as films on Fe-buffered MgO.
\end{abstract}


\maketitle

Since the discovery of superconductivity in K-doped BaFe$_2$As$_2$ (Ba-122) \cite{Rotter2008a}  many efforts have been made to investigate the physical properties and to understand the physics in these superconducting materials. Compared to single crystals, thin films are suitable for investigating transport and optical properties due to their dimensionality. Additionally, thin films are favourable for superconducting electronics applications. The symmetry of the superconducting order parameter has been investigated by means of hybrid Josephson junctions using Co-doped Ba-122 thin films realizing S'NS-type Josephson junctions as well as SQUID using bicrystal junctions.\cite{Katasesquid, ISI:000284233600037, 0953-2048-25-8-084020, 2012arXiv1211.3879S}\\
\indent Many groups have prepared thin films of the so called "122" family with \textit {AE}Fe$_2$As$_2$ (\textit {AE}: alkali earth elements, Sr and Ba) by means of pulsed laser deposition (PLD) and molecular beam epitaxy (MBE).\cite{leeKdopedfilms, TakedaMBESrBa, APEX.1.101702,HiramatsuWaterindsc, lee:212505, Katase20092121, iida:192501, Katase20092121, Adachi:2012:0953-2048:105015} We have proposed the usage of Fe as a buffer layer on a MgO (001) substrate for growing Co-doped Ba-122 thin films, which yields high crystalline quality as well as improved superconducting properties under our deposition conditions.\cite{Thersleff2010}

\begin{figure}[h]
\centering
\includegraphics[width=\columnwidth]{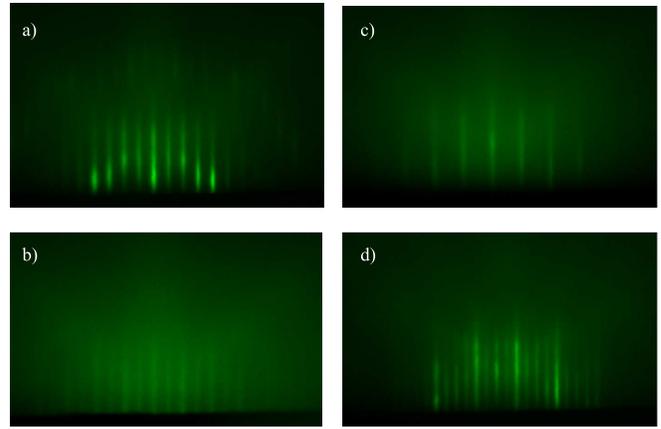}
\caption[RHEED images of the in-situ growth of the Ba-122 phases]{RHEED images of the $in\ situ$ growth of the Co-doped Ba-122 phase on (a) CaF$_2$, (b) SrF$_2$, (c) BaF$_2$ and (d) Fe-buffered MgO substrates. The pictures were taken after the deposition. The incident electron beam is along the substrates [110] azimuth. Since Ba-122 is grown with a 45\textdegree\ rotation on the fluoride substrates, the incident electron beam for (a), (b) and (c) is along the Ba-122 [100] azimuth. For (d) along the Ba-122 [110] azimuth.}
\label{RHEED_all}
\end{figure}

However, the Fe buffer has some disadvantages, such as the shunting of the current in the normal state which necessitates a recalculation of the superconducting transition temperature, $T_\mathrm{c}$.\cite{tro2012} Additionally, measurements of the normal state properties are always hindered by the Fe buffer layer.
For instance, the parent compound of Ba-122 has shown a spin density wave anomaly in $R(T)$ curves at around 140~K,\cite{Rotter2008} however, this transition is hardly recognized in the corresponding thin films on Fe-buffered MgO substrates. Furthermore, the intrinsic physical properties may be influenced by the ferromagnetism of the Fe or FeCo buffer layer, which is currently topic of investigation.\cite{italysalerno} Another significant influence of the Fe buffer is the observation of Co diffusion from the film into the Fe buffer layer, which leads to off-stoichiometry of the films.\cite{kurth1} In order to avoid such problems without compromising crystalline quality and superconducting properties, new types of substrates should be explored.\\ 
\indent Recently, epitaxial Fe(Se,Te) thin films as well as \textit {RE}FeAs(O,F) (\textit{RE}: Nd and Sm) thin films have been prepared on CaF$_2$ (001) substrates by PLD and MBE respectively.\cite{JJAP.51.010104,APEX.4.053101,Ueda2011,Uemura2012735} These films show high crystalline quality as well as excellent  superconducting properties. Hence, fluorides may be versatile substrates for Fe-based superconducting thin films. In this letter, we demonstrate the epitaxial growth of Co-doped Ba-122 on various fluoride substrates and discuss the influence of substrate materials on the structural and superconducting properties.\\ 
\indent For the deposition process a BaFe$_{1.84}$Co$_{0.16}$As$_2$ bulk target prepared by a solid-state-reaction was used ($T_\mathrm{c}$~= 23.7~K, measured by magnetization method). The phase purity of the target material was studied by powder X-ray diffraction (XRD) in Bragg-Brentano geometry (CoK$_{\alpha}$ radiation with $\lambda$ = 1.7889~\AA), showing high phase purity. The detailed preparation procedure and analysis can be found in Ref.\cite{kurth1}\\

\begin{figure}[htbp]
\centering
\includegraphics[width=\columnwidth]{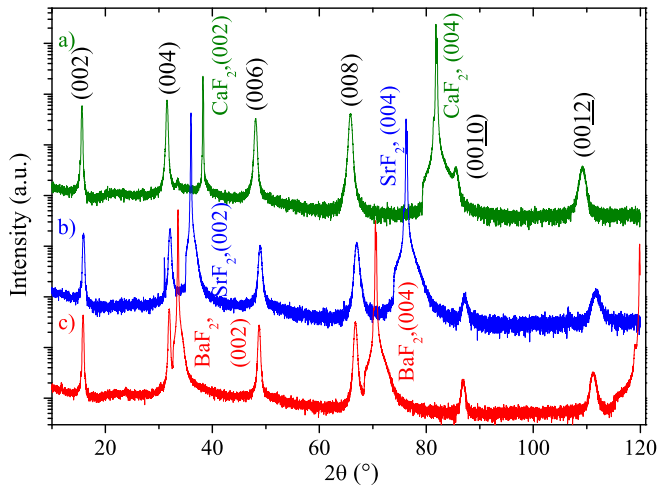}
\caption[theta - 2 theta scans for the Ba-122 thin films ] {$\theta$ - 2 $\theta$ scans for the Ba-122 thin films deposited on (a) CaF$_2$ (001), (b) SrF$_2$ (001) and (c) BaF$_2$ (001) substrates, respectively. A weak BaF$_2$ (002) reflection appears in the film deposited on CaF$_2$ and SrF$_2$.}
\label{xrd}
\end{figure}

The fluoride substrates were cleaned in an ultrasonic bath using acetone and isopropanol and subsequently transferred into the ultra high vacuum chamber (base pressure 10$^{-9}$~mbar). Co-doped Ba-122 layers were deposited on (001) \textit {AE}F$_2$ (\textit {AE}: Ca, Sr and Ba) single crystalline substrates using a KrF excimer laser ($\lambda$ = 248~nm, pulse duration = 25~ns) at a frequency of 7~Hz with an energy density of around 2.5~J/cm$^2$ at the target. The samples investigated in this study are summarized in Table \ref{tab:dep}.

\begin{table}[htbp]
\caption{Deposition temperature and  layer thickness of the Co-doped Ba-122 thin films.}
\label{tab:dep}
\centering
\begin{tabular}{ccc}\hline
{\textbf{Substrate}}&{\textbf{Deposition temp.}}&{\textbf{Layer thickness}}\\
\hline
CaF$_2$&750~\textcelsius&85~nm\\
CaF$_2$&700~\textcelsius&50~nm\\
SrF$_2$&700~\textcelsius&40~nm\\
BaF$_2$&700~\textcelsius&40~nm\\
\hline
\end{tabular}
\end{table}

\begin{figure}[htp]
\centering
\includegraphics[width=\columnwidth]{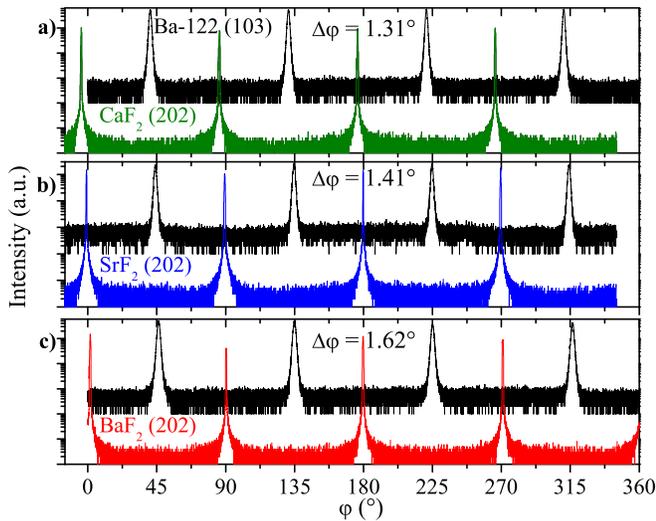}
\caption[Phiscans of the (103) refl of Ba-122 on BaF2 and CaF2] {$\phi$ scans of the (103) reflection of the Ba-122 phase on (a) CaF$_2$, (b) SrF$_2$ and (c) BaF$_2$ showing sharp reflections with a four-fold symmetry. No rotated grains are observed.}
\label{phiscan}
\end{figure}

The deposition process was monitored $in\ situ$ by reflection high energy electron diffraction (RHEED). In all cases streak like patterns were observed during film growth pointing to an epitaxial growth with low surface roughness. The spacing between streaks indicates the formation of a superstructure during the growth of the Co-doped Ba-122 on CaF$_2$ and SrF$_2$, i.e. a surface reconstruction that is also observed in single crystals (Fig.\ref{RHEED_all}(a), (b)).\cite{PhysRevLett.106.027002} These RHEED patterns are very similar to Co-doped Ba-122 thin films on Fe-buffered MgO (Fig.\ref{RHEED_all}(d)).\cite{Iida:2011:0953-2048:125009} For Co-doped Ba-122 on BaF$_2$, despite the observation of streaks, however, no sign of this superstructure is observed, presumably due to a relatively large misfit of -10.6\%. Therefore we assume a lower surface quality for the film grown on BaF$_2$.\\
\indent The phase purity of the thin films was confirmed by X-ray diffraction (Fig.~\ref{xrd}). The scans showed no additional reflections of other phases except for the films deposited on both, CaF$_2$ and SrF$_2$, where a small peak at 2$\theta$ = 33.6\textdegree\ is visible, which can be assigned to the (002) plane of BaF$_2$. The origin of this reflection will be discussed below.\\
\indent The determination of the epitaxial relationship between films and substrates was conducted by XRD texture measurements. The (103) and (202) planes have been used for Co-doped Ba-122 and fluoride substrates, respectively. All films showed a clear fourfold symmetry without any additional peaks. The film reflections are rotated 45\textdegree\ with respect to the substrate. Accordingly the films are grown epitaxially with the relation (001)[110]Ba-122$||$(001)[100]\textit{AE}F$_2$. The corresponding $\phi$ scan of Co-doped Ba-122 and substrates are displayed in Fig. \ref{phiscan}. The average full width at half maximum ($\Delta\phi$) values of films grown on CaF$_2$, SrF$_2$ and BaF$_2$ are 1.31\textdegree, 1.41\textdegree\ and 1.62\textdegree, respectively. Values were not corrected for device broadening. We assume that the large lattice misfit (Table \ref{tab:misfit}) leads to a relatively large $\Delta\phi$\ value. Measurements on a thicker Co-doped Ba-122 film (85~nm) on CaF$_2$ showed an even higher crystalline quality with a $\Delta\phi$ value of 0.95\textdegree, which is almost the same value of layers grown on Fe-buffered MgO.\cite{Iida:2011:0953-2048:125009}

\begin{table}[htbp]
\caption{Lattice parameters and misfit data of the Ba-122 on different substrates.}
\label{tab:misfit}
\centering
\begin{tabular}{ccccc}\hline
\multirow{1}{*}{\textbf{Substrate}}&\multicolumn{1}{c}{\textbf{Substrate}}&\multicolumn{2}{c}{\textbf{Thin films}}&{Misfit in}\\
\cline{2-4}
Ba-122 thickness&a-axis/$\sqrt{2}$&a-axis&c-axis&a-axis\\
\hline
\textbf{CaF$_2$} - 85~nm&3.854~\AA&3.89~\AA&13.19~\AA&+2.7\%\\
\textbf{CaF$_2$} - 50~nm&3.854~\AA&3.91~\AA&13.17~\AA&+2.7\%\\
\textbf{SrF$_2$} - 40~nm&4.100~\AA&3.94~\AA&12.97~\AA&-3.2\%\\
\textbf{BaF$_2$} - 40~nm&4.381~\AA&3.96~\AA&13.02~\AA&-10.6\%\\
\hline
\end{tabular}
\end{table}

As stated earlier, Co-doped Ba-122 films have been grown rotated 45\textdegree\ in-plane on the substrates. Therefore, the lattice misfit between film and substrate is defined as $((a_{Ba-122}\times\sqrt{2})-a_{substrate})/(a_{Ba-122}\times\sqrt{2})$.
In this calculation, the lattice parameter $\textit{a}$ = 3.96~\AA\ of bulk (i.e. PLD target) was used. In Table \ref{tab:misfit}, the in-plane and out-of-plane lattice parameters of Co-doped Ba-122 evaluated through high resolution reciprocal space maps (RSM) are also summarized. The out-of-plane lattice parameter evaluated by the Nelson-Riley extrapolation from $\theta-2\theta$ scans are in good agreement with the values of the RSM measurements. Clearly, negative misfit values of films on SrF$_2$ and BaF$_2$ lead to a slightly larger in-plane lattice parameter compared to the one on CaF$_2$. However, a change in lattice parameter \textit{a} is small even for large misfit, in particular, BaF$_2$ substrate, whereas a significant change in lattice parameter \textit{c} is observed. Additionally, a reaction layer between Co-doped Ba-122 and substrates is observed, which will be discussed later. Hence, such a change in lattice parameters can not be explained by the simple misfit scenario indicative of another mechanism such as F intercalation into the Co-doped Ba-122 lattice. Possible diffusion of F from the fluoride substrate has been discussed in both the SmFeAs(O,F) and the Fe(Se,Te) systems.\cite{Takeda:2012:0953-2048:35007,2012arXiv1208.3287I} Further investigations are underway.

\begin{figure}[htbp]
\centering
\includegraphics[width=\columnwidth]{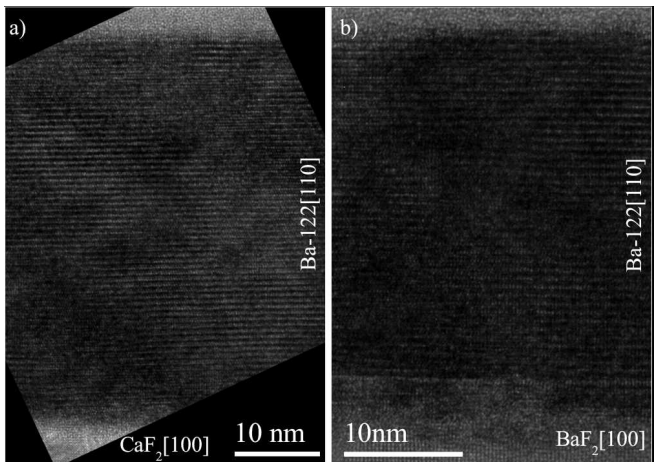}
\caption[TEM pictures of BaF2 and CaF2 thin films] {TEM bright field monograph of the (a) Ba-122/CaF$_2$ (thickness about 50\ nm) and the (b) Ba-122/BaF$_2$ thin film (thickness about 40\ nm) showing high structural quality of the Ba-122 layer. However a reaction at the interface is observed.}
\label{tem_baf2}
\end{figure}

\begin{figure}[htbp]
\centering
\includegraphics[width=\columnwidth]{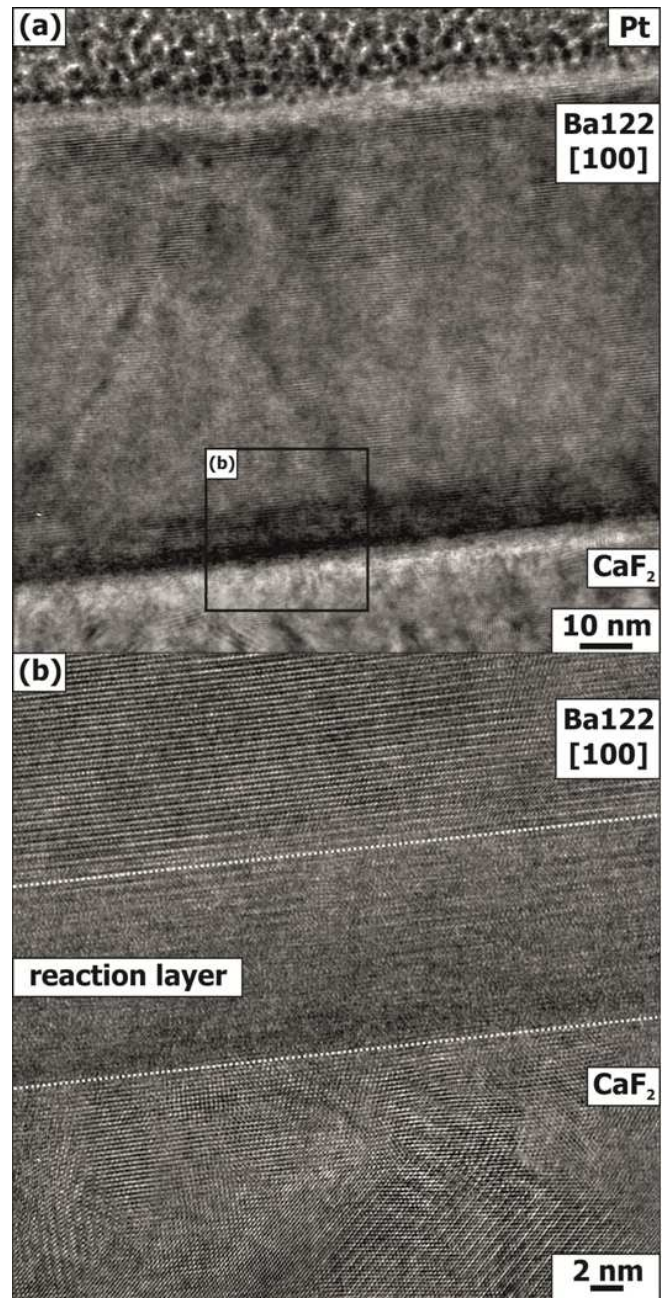}
\caption[HRTEM pictures of BaF2 and CaF2 thin films] {HRTEM investigations of the Ba-122 film (thickness about 85~nm) deposited on CaF$_2$ (a). A closer look at the Ba-122-CaF$_2$-substrate interface (b) reveals a reaction layer which can be assumed to be BaF$_2$ due to XRD and EELS data.}
\label{tem_caf2}
\end{figure}

TEM investigations of the films grown on CaF$_2$ and BaF$_2$ showed no appreciable defects in the Co-doped Ba-122 layers (Figs. \ref{tem_baf2}). However a reaction at the interface is observed. The respective film thickness on CaF$_2$ and BaF$_2$ substrates are 40~nm and 50~nm. For the film on SrF$_2$, a layer thickness of 40~nm was determined by a cross-sectional focused ion cut on the sample.\\

\begin{figure}[ht]
\centering
\includegraphics[width=\columnwidth]{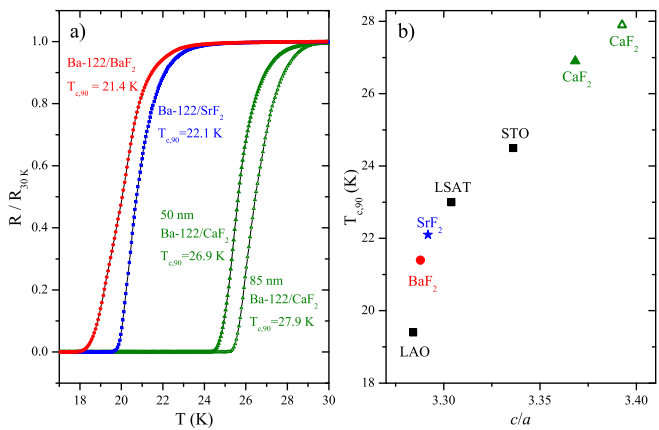}
\caption[resistivity measurements which shows the sc transition] {(a) Resistance measured curves for the Co-doped Ba-122 deposited upon BaF$_2$ (red round), SrF$_2$ (blue star) and CaF$_2$ (green triangle). {\itshape T$_\mathrm{c,90}$} is 21.4~K for the film deposited on BaF$_2$, 22.1~K for the film on SrF$_2$, 26.9~K on CaF$_2$ and 27.9~K for the thicker film on CaF$_2$ (green half-filled triangle). (b) T$_{c,90}$ over \textit{c/a} shows almost linear dependence. The film with the highest \textit{c/a} ratio shows the highest {\itshape T$_\mathrm{c,90}$} (green half-filled triangle). The data from oxide substrates are also plotted on the same figure.\cite{iida:192501}}
\label{rt}
\end{figure}

Microstructual investigations on a thicker film (i.e. better crystalline quality) by high resolution TEM (HRTEM), high resolution scanning TEM (HRSTEM), and electron energy loss spectroscopy (EELS) have also been conducted. Although a possible fluorine diffusion into the Co-doped Ba-122 has been discussed earlier, a trace of fluorine in the layer cannot be detected by EELS due presumably to the high volatility of the fluorine under the influence of the electron beam. An interfacial layer ($\sim$10~nm) between the Co-doped Ba-122 and the CaF$_2$ substrate is clearly observed (Figs. \ref{tem_caf2}).
A HRSTEM linescan over the interface with combined Ba detection using EELS revealed a small diffusion of Ba into the CaF$_2$ substrate. As stated earlier, a small BaF$_2$ signal was detected in $\theta$-2$\theta$ scans (Fig. \ref{xrd}). Accordingly, we assume this reaction layer to be BaF$_2$.

The $T_\mathrm{c,90}$, which is defined as 90~\% of the resistance in the normal state at 30~K, of the thin films was determined through resistance measurements with a four point probe method using a Physical Properties Measurement System (Fig. \ref{rt}(a)). The film deposited on BaF$_2$ shows a $T_\mathrm{c,90}$ of 21.4~K and the film on SrF$_2$ shows 22.1~K whereby the film deposited on CaF$_2$ has the highest value of 26.9~K. It is worth mentioning that the highest $T_\mathrm{c}$ of over 28~K has been realized in thick Co-doped Ba-122 on CaF$_2$. Additionally, this film shows a high self-field critical current density ($J_\mathrm{c}$) of over 1~MA/cm$^2$ even at 10~K and good in-field $J_\mathrm{c}$ performances (not shown in this letter).
Hence fluoride substrates, in particular CaF$_2$, offer a clear benefit for growing epitaxial Co-doped Ba-122 with good crystalline quality and superconducting properties.   
Fig. \ref{rt}(b) shows the relationship between $T_\mathrm{c}$ and the \textit{c/a} ratio. For comparison, the films on various oxide substrates are also plotted in the same figure.\cite{iida:192501} It is clear from fig. \ref{rt}(b) that $T_\mathrm{c}$ is increasing with increasing \textit{c/a}. From the above results we conclude that the lattice distortion clearly affects the superconducting properties. The origin of the trend of the $T_\mathrm{c,90}$(\textit{c/a}) needs further investigations.

Here we emphasize that the crystalline quality of the thicker films grown on CaF$_2$ is almost identical if not superior to the films on Fe-buffered MgO. Additionally, several advantages over the use of Fe buffer layers are highly expected such as the absence of current shunting due to the Fe layer.\cite{tro2012} Most importantly, the stoichiometry of the Co-doped Ba-122 film, in particular, the Co concentration, can be more stable, since Co diffusion into the Fe-buffer is avoidable.\\

To conclude, Co-doped Ba-122 thin films have been grown on various (001) fluoride substrates by PLD. Despite the presence of a possible BaF$_2$ reaction layer, all films have been grown epitaxially with good crystalline quality. In particular, the crystalline quality of the Co-doped Ba-122 on CaF$_2$ substrate is almost identical if not superior to that on Fe-buffered MgO substrate. The lattice distortion of the Co-doped Ba-122, which influences the $T_\mathrm{c}$ significantly, is confirmed by RSM. The highest $T_\mathrm{c}$ of about 28~K has been achieved on CaF$_2$. Therefore the CaF$_2$ substrate offers a clear benefit for growing high-quality, epitaxial Fe-based superconducting thin films.\\

The authors thank S.\,F\"ahler and V.\,Grinenko for valuable discussions as well as U.\,Besold, J.\,Scheiter, M.\,Langer and M.\,K\"uhnel for technical support. The research leading to these results has received funding from European Union's Seventh Framework Programme (FP7/2007-2013) under grant agreement number 283141 (IRON-SEA) and by Strategic International Collaborative Research Program (SICORP), Japan Science and Technology Agency.


%

\end{document}